\documentclass[11pt]{article}
\usepackage{epsfig}
\usepackage{bm}
\usepackage{amssymb}
\usepackage{mathrsfs}
\usepackage{amsmath}
\usepackage{dsfont}

\begin{document}
\thispagestyle{empty}
$\;$ \hfill INT-PUB-12-013
\vspace*{10mm}

\begin{center}

{\LARGE 
Monte Carlo simulation of the SU(3) spin model \vskip1mm
with chemical potential in a flux representation}
\vskip15mm
Ydalia Delgado Mercado$^{\, a,b}$, 
Christof Gattringer$^{\, a,c}$  
\vskip5mm
$^a\,$Karl-Franzens University Graz \\
Institute for Physics\\ 
Universit\"atsplatz 5, A-8010 Graz, Austria 
\vskip5mm
$^b\,$Bergische Universit\"at Wuppertal \\
Department of Physics \\
Gau{\ss}str.~20, D-42119 Wuppertal, Germany
\vskip5mm
$^c\,$University of Washington, Seattle \\
Institute for Nuclear Theory \\
Box 351560, Seattle, WA 98195, USA
\end{center}
\vskip15mm

\begin{abstract}
We present a simulation of the SU(3) spin model with chemical potential using a
recently proposed flux representation. In this representation the complex phase problem
is avoided and a Monte Carlo simulation in terms of the fluxes becomes possible. We
explore the phase diagram of the model as a function of temperature and chemical
potential.
\end{abstract}

\vskip20mm
\noindent
{\tt ydalia.delgado-mercado@uni-graz.at \\
christof.gattringer@uni-graz.at}

\setcounter{page}0
\newpage
\noindent
\section{Introductory remarks}
The complex phase problem (fermion sign problem) is the central obstacle that has held
up Monte Carlo simulations of finite density lattice QCD for two decades. When a
chemical potential is coupled the fermion determinant becomes complex and cannot be
directly used as a probability weight in a Monte Carlo simulation. The alternative
approaches that were explored, such as reweighting, power series expansion, strong
coupling/large mass expansion or analytic continuation from imaginary chemical have
had only limited success so far. 

For several systems which are simpler than QCD the complex phase problem was solved 
by rewriting the theory in terms of new degrees of freedom \cite{others,gattringer} 
where the complex phase problem is avoided and simulations with new techniques such as worm
algorithms  \cite{worm} become possible. Several of these models were also explored
with another method for systems with sign problems, the complex Langevin
approach \cite{wyldkarsch,gausterer,langevin,aarts}. 

In this article we present results for the SU(3) spin
model \cite{wyldkarsch}, which at vanishing external magnetic field is expected to 
describe the deconfinement
transition \cite{center} of pure SU(3) gauge theory. Adding the magnetic field terms takes
into account the leading corrections from the fermion determinant and couples the
chemical potential (see \cite{strong} for a treatment of
the model beyond these leading contributions). The model has been studied successfully
using the complex Langevin approach \cite{wyldkarsch,gausterer,aarts} and has a flux representation
\cite{gattringer} where the complex phase problem is avoided. In this paper we discuss 
a Monte Carlo simulation of the SU(3) spin model based on the flux 
representation and explore the phase diagram of the model as a function of temperature
and chemical potential. 

\section{The model and its flux representation}
The action of the SU(3) spin model is given by 
\begin{equation}
S \;  = \; - \!\sum_x \left(\! \tau \! \sum_{\nu = 1}^3 \! \Big[ P(x) P(x+\hat{\nu})^\star
+ c.c. \Big] 
+ \kappa \Big[ e^\mu P(x) +  e^{-\mu} P(x)^\star \Big]\! \right)\; .
\label{action_su3} 
\end{equation}
The degrees of freedom $P(x)$, which we sometimes will  refer to as Polyakov loops,
are the traced SU(3) variables $P(x) =$ Tr $L(x)$ with $L(x) \in$ SU(3). They are
attached to the sites $x$  of a three-dimensional cubic lattice which we consider to
be finite with periodic boundary conditions. By $\hat{\nu}$ we denote the unit vector
in $\nu$-direction, with $\nu = 1, 2, 3$. The first two terms are a nearest
neighbor interaction between adjacent Polyakov loops. The parameter $\tau$ depends on
the temperature (it increases with temperature) and is real and positive. The real and
positive parameter $\kappa$ is proportional to the number of flavors and depends on
the fermion mass (it decreases with $m_q$) and $\mu$ is the chemical potential. For
later use we introduce the abbreviations $\eta = \kappa e^{\mu}$ and 
$\overline{\eta} = \kappa e^{-\mu}$.

The grand canonical partition function of the model described by (\ref{action_su3}) is
obtained by integrating the Boltzmann factor $e^{-S[L]}$ over all configurations of
the Polyakov loop variables. The corresponding measure is a product over the reduced
Haar measures $dP(x)$ at the sites $x$. Thus

\begin{equation}
Z \; = \int_{SU(3)} \! \prod_x dL(x) \, e^{-S[L]} \; .
\label{sum_su3} 
\end{equation}

Without the magnetic term, i.e., for $\kappa = 0$, the system has a low 
temperature phase where the expectation value $\langle P \rangle$ for the spatially 
summed Polyakov loop $P = \sum_x P(x)$ vanishes which is interpreted as the confined phase of
QCD. At $\tau_c \sim 0.137$ the system undergoes a first order deconfinement transition
into a phase which is characterized by $\langle P \rangle \neq 0$ (deconfined phase). 
For small $\kappa$ and $\mu = 0$ the first order line persists ending in a second
order endpoint \cite{wyldkarsch}. We will show here that for $\mu > 0$ the first order 
transition is weakened further, i.e., the endpoint shifts towards smaller $\kappa$.  

Applying high temperature expansion techniques, the partition function can be
rewritten in terms of new degrees of freedom, the flux variables. The general steps
to obtain the flux representation are \cite{gattringer}:

\begin{itemize}
\item The Boltzmann factor is written as a product over all nearest neighbor terms
and a product over all sites for the magnetic terms, respectively. Each individual
exponential is then expanded in a power series.

\item
The emerging products can be reorganized, such that at each lattice site $x$ one has a
(reduced) Haar measure integral over powers of $P(x)$ and $P(x)^\star$. The exponents 
are combinations of the expansion variables of the individual exponential terms of the
Boltzmann factors.

\item
Based on techniques presented in \cite{wipfsu3} these integrals 
(moments of the one-link integrals) can be solved in closed form. 
The integrals give rise to local constraints for the 
allowed combinations of the expansion coefficients of the exponentials.
\end{itemize}

Obviously these steps are a straightforward application of textbook high temperature
expansion techniques applied to a somewhat unusual spin system. The final result 
for the flux representation of the partition sum is given by \cite{gattringer}:
\begin{equation}
Z \; = \; \sum_{\{l,\overline{l}\}} \sum_{\{s,\overline{s}\}}
 \left( \prod_{\overline{x},\nu} 
 \frac{\tau^{l_{x,\nu}+\overline{l}_{x,\nu}}}{l_{x,\nu}! \; \overline{l}_{x,\nu}!}  \right) 
 \left( \prod_x \frac{\eta^{s_x} \; \overline{\eta}^{\overline{s}_x}}{s_x! \; \overline{s}_x!} \right) 
 \left( \prod_x  I(f_x,\overline{f}_x) \right) \; .
 \label{zflux}
\end{equation}
In this form the partition sum is a sum over all configurations of the two sets of 
dimer variables $l_{x,\nu}, \overline{l}_{x,\nu} \in [0,+\infty)$ , 
living on the links $(x,\nu)$ and the monomers  $s_x,\overline{s}_x \in 
[0,+\infty)$, living on the sites $x$. The first two products in (\ref{zflux})
are weight factors that come from the expansion of the individual exponentials 
and it is obvious that in the second product the chemical potential $\mu$ 
enters via the powers of $\eta = \kappa e^{\mu}$ and 
$\overline{\eta} = \kappa e^{-\mu}$.
 
The last product is over the group integrals at each lattice site $x$, 
\begin{equation}
I(n,\overline{n}) \; = \; \int_{SU(3)} \!\!\!\! dL \, (\mbox{Tr} L)^{n} \,
(\mbox{Tr} L^\dagger)^{\overline{n}} \; , 
\end{equation}
where $dP$ denotes SU(3) Haar measure and 
$n$ and $\overline{n}$ are non-negative integers. These integrals 
can be evaluated in closed form \cite{gattringer} and turn out to be real and
non-negative. However, the $I(n,\overline{n})$ 
are non-zero only if the triality condition 
$(n - \overline{n} ) \, \mbox{mod} \, 3 = 0$ is obeyed. 
In the flux representation (\ref{zflux}) for the partition sum the arguments 
of the $I$ are the summed fluxes $f_x$ and $\overline{f}_x$ 
at the sites $x$ of the lattice defined by 
\begin{equation}
f_x \; = \; \sum_{\nu=1}^{3} [\, l_{x,\nu}+\overline{l}_{x-\hat{\nu},\nu} \, ]  +  s_x 
\quad, \qquad
\overline{f}_x \; = \; \sum_{\nu=1}^{3}
[\, \overline{l}_{x,\nu} + l_{x-\hat{\nu},\nu} \, ] + \overline{s}_x \; .
\end{equation}
The triality condition for non-vanishing weights $I$ then reads
\begin{equation}
( \, f_x \, - \, \overline{f}_x \, ) \, \mbox{mod} \, 3 \; = \; 0 \; ,
\label{constraint}
\end{equation}
which introduces a constraint for the allowed values of the dimer and monomer 
variables at each lattice site $x$. 

Thus in the form (\ref{zflux}) the partition sum is a sum over configurations of the
dimers $l_{x,\nu}, \overline{l}_{x,\nu} \in [0,+\infty)$ , 
and the monomers  $s_x,\overline{s}_x \in 
[0,+\infty)$ with weight factors that are real and non-negative. For a successful
solution to the complex phase problem one still has to find a Monte Carlo  
update which only generates configurations that obey the constraint
(\ref{constraint}). Otherwise the sample will be dominated by configurations with zero
weight, and the signal will vanish exponentially with the volume.

\section{Reparametrization and observables}
A first version of our Monte Carlo update of the flux representation was set up
directly in terms of the flux representation given in (\ref{zflux}). It turned out
that this algorithm did not perform very well, and we found that a reparametrization 
of the dimer and monomer variables gives rise to an alternative flux representation which
allows for a considerably better algorithm.

We introduce new integer valued dimer variables 
$k_{x,\nu} \in \ [0,+\infty)$ and $\overline{k}_{x,\nu} \in \ (-\infty,+\infty)$, which are
related to the old dimer variables via $l_{x,\nu} - \overline{l}_{x,\nu} = \overline{k}_{x,\nu}$
and $l_{x,\nu} + \overline{l}_{x,\nu} = |\overline{k}_{x,\nu}| + 2k_{x,\nu}$. The various terms
in the partition sum change according to
\begin{itemize}
\item[] $\tau^{l_{x,\nu}+\overline{l}_{x,\nu}} \; \rightarrow \;
\tau^{|\overline{k}_{x,\nu}| + 2k_{x,\nu}} \; , $
\vspace{-1mm}
\item[] $l_{x,\nu}! \, \overline{l}_{x,\nu}! \; \rightarrow \; 
(|\overline{k}_{x,\nu}| + k_{x,\nu})! \, k_{x,\nu}! \; . $ 
\end{itemize}

Similarly we define new integer valued monomer variables $r_x \in \ [0,+\infty)$ and 
$\overline{r}_x \in \ (-\infty,+\infty)$ which are related to the old monomer variables via
$s_x - \overline{s}_x = \overline{r}_x$ and $s_x + \overline{s}_x = |\overline{r}_x| + 2r_x$.
The corresponding replacements in the partition sum 
are:
\begin{itemize}
\item[] $\kappa^{s_x+\overline{s}_x} \; \rightarrow \; \kappa^{|\overline{r}_x| + 2r_x} \; ,$
\vspace{-1mm}
\item[] $e^{\mu (s_x-\overline{s}_x)} \; \rightarrow \; e^{\,\mu \,\overline{r}_x} \; ,$
\vspace{-1mm}
\item[] $s_x! \,\overline{s}_x! \; \rightarrow \; (|\overline{r}_x| + r_x)!\, r_x! \; \; .$ 
\end{itemize}
Also the fluxes $f_x$ and $\overline{f}_x$ may be rewritten in terms of 
the new dimer and monomer variables  
\begin{eqnarray}
f_x &\!\!\! =\!\!\! & \sum_{\nu} \! \left[\frac{|\overline{k}_{x,\nu}|\!+\!|\overline{k}_{x-\nu,\nu}|}{2} 
+ k_{x,\nu} + k_{x-\nu,\nu} + \frac{\overline{k}_{x,\nu} \!-\! \overline{k}_{x-\nu,\nu }}{2}\right] + 
\frac{|\overline{r}_x| \!+\! \overline{r}_x}{2} + r_x \; ,
\nonumber \\
\overline{f}_x &\!\!\! =\!\!\! & \sum_{\nu} \! \left[\frac{|\overline{k}_{x,\nu}|\!+\! |\overline{k}_{x-\nu,\nu}|}{2} 
+ 
k_{x,\nu} + k_{x-\nu,\nu} - \frac{\overline{k}_{x,\nu} \!-\! \overline{k}_{x-\nu,\nu}}{2}\right] + 
\frac{|\overline{r}_x| \!-\!\overline{r}_x}{2} + r_x \; .
\nonumber \\
\label{newflux}
\end{eqnarray}  
After the reparametrization the partition function thus reads:
\begin{equation}
Z = \sum_{\{k,\overline{k}\}} \sum_{\{r,\overline{r}\}} \!
\left( \prod_{x,\nu} \frac{\tau^{|\overline{k}_{x,\nu}| + 
2k_{x,\nu}}}{(|\overline{k}_{x,\nu}| \!+\! k_{x,\nu})! \, k_{x,\nu}!} \right)\!\!
\left( \prod_x \frac{\kappa^{|\overline{r}_x| + 2r_x} \,
e^{\mu \overline{r}_x} }{(|\overline{r}_x| \!+\! r_x)! \, r_x!} \right)\!\!
\left( \prod_x I(f_x,\overline{f}_x) \right) ,
\end{equation}
where the sum is now over all configurations of the new dimer variables 
$k_{x,\nu} \in \ [0,+\infty)$ and $\overline{k}_{x,\nu} \in (-\infty,+\infty)$
and the new monomer variables $r_x \in [0,+\infty)$ and 
$\overline{r}_x \in (-\infty,+\infty)$. For the fluxes 
$f_x$ and $\overline{f}_x$ now the expressions (\ref{newflux}) are used. 

It is interesting to note that in terms of the new variables the triality constraint 
$( f_x  -  \overline{f}_x ) \, \mbox{mod} \, 3 \, = \, 0$ from Eq.~(\ref{constraint})
turns into
\begin{equation}
\Big(\sum_{\nu}[\overline{k}_{x,\nu} - \overline{k}_{x-\nu,\nu}] + \overline{r}_x \Big) \!
\mod 3 \; = \; 0 \; .
\label{newconstraint}
\end{equation}
Obviously the reparametrization simplified the constraint and only the variables
$\overline{k}_{x,\nu}$ and $\overline{r}_x$ enter the constraint. The variables 
$k_{x,\nu}$ and $r_x$ can be varied freely without generating zero weight configurations.
We stress that also the chemical potential only couples to the $\overline{r}_x$ monomers. 
We attribute the aforementioned better performance of the Monte Carlo simulation after the 
reparametrization to both these properties: The fact that the constraint restricts only the
bared variables $\overline{k}_{x,\nu}$ and $\overline{r}_x$ and that the chemical potential
couples only to the $\overline{r}_x$.

The observables we consider are obtained as derivatives of $\ln Z$ with respect to the physical 
parameters $\tau, \mu$ and $\kappa$, and also with respect to the combined parameters
$\eta = \kappa e^\mu$ and $\overline{\eta} = \kappa e^{-\mu}$. Defining the following abbreviations
for the sums of dimer and monomer variables,
\begin{equation}
K \equiv \sum_{x,\nu} [ |\overline{k}_{x+\nu}| + 2k_{x+\nu} ] \; , \;  
R \equiv \sum_x r_x  \; , \;
\overline{R} \equiv \sum_x \overline{r}_x \; , \;
|\overline{R}| \equiv \sum_x |\overline{r}_x| \; ,
\end{equation}
one easily checks the following auxiliary identities for the first and second derivatives of the
partition sum in the flux representation
\begin{eqnarray}
&& 
\frac{1}{Z}\frac{\partial}{\partial \tau}Z = 
\frac{1}{\tau}\Big\langle K \Big\rangle \; ,
\nonumber \\
&& 
\frac{1}{Z}\frac{\partial}{\partial \eta}Z = 
\frac{1}{\eta}\left\langle \frac{|\overline{R}|+\overline{R}}{2} + R \right\rangle \; ,
\nonumber \\
&&
\frac{1}{Z}\frac{\partial}{\partial \overline{\eta}}Z = 
\frac{1}{\overline{\eta}}\left\langle 
\frac{|\overline{R}|-\overline{R}}{2} + R \right\rangle \; ,
\nonumber \\
&&
\frac{1}{Z}\frac{\partial^2}{\partial \tau^2}Z = 
\frac{1}{\tau^2}\Big\langle K^2 - K \Big\rangle \; ,
\nonumber \\
&&
\frac{1}{Z}\frac{\partial^2}{\partial \eta^2}Z = 
\frac{1}{\eta^2}\left\langle 
\left( \frac{|\overline{R}|+\overline{R}}{2} + R \right)^2 - 
\left(\frac{|\overline{R}|+\overline{R}}{2} + R\right) \right\rangle \; ,
\nonumber \\
&&
\frac{1}{Z}\frac{\partial^2}{\partial \overline{\eta}^2}Z = 
\frac{1}{\overline{\eta}^2}\left\langle \left( \frac{|\overline{R}|-\overline{R}}{2} + R \right)^2 
- \left(\frac{|\overline{R}|-\overline{R}}{2} + R\right) \right\rangle \; .
\label{auxident}
\end{eqnarray}
From the original form (\ref{action_su3}) of the action one may identify the physical interpretation 
of the observables obtained by derivatives of $\ln Z$. Using the auxiliary identities
(\ref{auxident}), one obtains the following flux representations for the internal energy $U$, the heat
capacity $C$, the Polyakov loop expectation value $\langle P \rangle$, and the Polyakov loop
susceptibility $\chi_P$,
\begin{eqnarray}
U &\!\!\!\!\!=\!\!\!\!\!& \frac{1}{Z}\!\!\left[\tau\frac{\partial}{\partial \tau} + \eta\frac{\partial}{\partial \eta}  + 
\overline{\eta}\frac{\partial}{\partial \overline{\eta}} \right]Z 
  = \left\langle\  K + |\overline{R}| + 2R\  \right\rangle \; ,
\nonumber \\
C &\!\!\!\!\!=\!\!\!\!\!& \frac{1}{Z}\!\!\left[\tau^2\frac{\partial^2}{\partial \tau^2} 
\!+\! \eta^2\frac{\partial^2}{\partial \eta^2}  \!+\! 
\overline{\eta}^2\frac{\partial^2}{\partial \overline{\eta}^2}
\!+\!2\tau\eta\frac{\partial^2}{\partial \tau \partial \eta}\!+\! 2\eta\overline{\eta}\frac{\partial^2}{\partial 
\eta \partial \overline{\eta}} \!+\! 
2\tau \overline{\eta}\frac{\partial^2}{\partial \tau \partial \overline{\eta}} \right]\!\!Z - U^2 \nonumber \\
&\!\!\!\!\!=\!\!\!\!\!& \left\langle\  \left[ \left(K + |\overline{R}| + 2R\right) - U\right]^2 - \left(K + |\overline{R}| + 2R\right)\  
\right\rangle \; ,
\nonumber \\
\langle P \rangle &\!\!\!\!\!=\!\!\!\!\!& \frac{1}{Z}\frac{\partial}{\partial \eta}Z \nonumber 
= \frac{1}{\eta} \left\langle\  \frac{|\overline{R}| + \overline{R}}{2} + R \right\rangle \; ,
\nonumber \\
\chi_P &\!\!\!\!\!=\!\!\!\!\!& \frac{1}{Z}\frac{\partial^2}{\partial \eta^2}Z - \langle P \rangle^2 
= \left\langle\ \!\! \left[ \frac{1}{\eta}\!\!\left(\frac{|\overline{R}| \!+\! \overline{R}}{2} + R\!\right) - 
\langle P \rangle \right]^2\!\! - \frac{1}{\eta^2}\!\!\left(\frac{|\overline{R}| \!+\! \overline{R}}{2}
+ R\!\right)\!\! \right\rangle .
\nonumber \\
\end{eqnarray}
Obviously all our observables can be computed as expectation values of sums of dimers and
monomers and their second moments. 
 
\section{MC algorithm and setup of the simulation}

After the reparametrization in the previous section the partition sum is a sum over configurations
of two classes of variables: The ''unbared'' dimer and monomer variables 
$k_{x,\nu}, r_x \in \ [0,+\infty)$ and the ''bared'' dimers and monomers,
$\overline{k}_{x,\nu}, \, \overline{r}_x  \in (-\infty,+\infty)$. 

The unbared variables $k_{x,\nu}, r_x$ are not subject to any constraint and we simply update
them by randomly raising or lowering them by 1 and accepting this change with the usual Metropolis 
probability. Applying this step once to all $k_{x,\nu}$ constitutes one sweep for the unbared dimers
and similarly one sweep for the unbared monomers is to run through all $r_x$.

The bared variables $\overline{k}_{x,\nu}, \, \overline{r}_x$ must obey the constraint 
(\ref{newconstraint}). The constraint forces the combined flux of bared dimer and monomer 
variables at every lattice point to be a multiple of 3. A starting configuration that obeys
the constraint is given by setting all bared dimers and monomers to zero. 
The following 5 update steps 
leave the constraints intact and can be seen to give rise to an ergodic update. Each individual 
step is accepted with the Metropolis probability. 

\begin{enumerate}

\item
The value of a dimer variable $\overline{k}_{x,\nu}$ is randomly changed by $\pm 3$. 
For the sites $x$ and $x+\widehat{\nu}$ the right hand side of (\ref{newconstraint}) 
changes by three units and thus the constraints at the two sites remain intact. Again full sweeps
through all $\overline{k}_{x,\nu}$ are implemented.

\item
The value of a monomer variable $\overline{r}_x$ is randomly changed by $\pm 3$. 
For the site $x$ the right hand side of (\ref{newconstraint}) 
changes by three units and thus the constraint at $x$ remains intact. Full sweeps through all 
$\overline{r}_x$ are done. 

\item
Along an oriented plaquette all dimer variables $\overline{k}_{x,\nu}$ are changed by
$\pm 1$ according to their orientation in the plaquette. For all
corners of the plaquette the left hand side of (\ref{newconstraint}) remains unchanged. A sweep
is defined as offering this step to all plaquettes on the lattice.

\item
The value of a $\overline{k}_{x,\nu}$ is randomly changed by $\pm 1$ and the monomers at its 
endpoints are changed accordingly: $\overline{r}_x \rightarrow \overline{r}_x \mp 1$, 
$\overline{r}_{x+\widehat{\nu}} \rightarrow \overline{r}_{x+\widehat{\nu}} \pm 1$. The left hand
sides of (\ref{newconstraint}) remain unchanged at the two endpoints of the dimer. Again we define a
sweep by offering this change to all bared dimers and the monomers at their endpoints.

\item
All values of dimer variables $\overline{k}_{x,\nu}$ that sit on a straight loop which closes around 
the periodic boundary are randomly changed by $\pm 1$ (same value for all of them). For all sites on
the loop the constraint remains unchanged. A sweep is defined as offering this change to all loops
in the 3 directions. At $\kappa = 0$ this step is necessary for ergodicity. 

\end{enumerate}   
A combined full update sweep consists of a sweep for the unbared dimers, a sweep 
for the unbared monomers, and one of each of the 5 update sweeps for the bared variables.

Our simulations were done on lattices with sizes between $6^3$ and $20^3$. We typically used
statistics of up to $10^6$ configurations, separated by 10 full update sweeps for decorrelation and 
$10^6$ sweeps for equilibration. All errors we quote are statistical errors determined with the 
Jackknife method. 

\section{Comparison to other approaches}

\begin{figure}[t!]
\centering
\hspace*{-12mm}
\includegraphics[width=1.16\textwidth,clip]{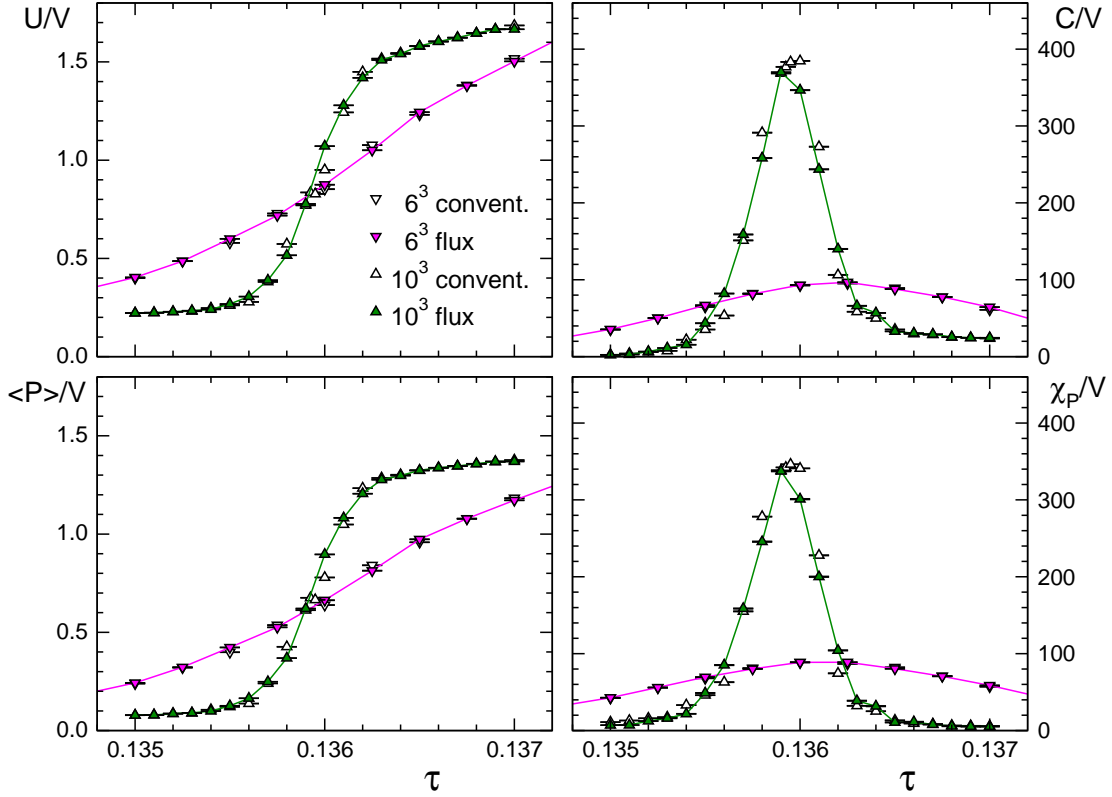}
\vspace*{-3mm}
\caption{Comparison of results from a simulation in the flux representation
(filled symbols) to data from a simulation in the conventional approach (empty
symbols). We show the results for $U, C, \langle P \rangle$ and $\chi_P$ 
as a function of $\tau$ at $\kappa = 0.005$ and $\mu = 0$ and compare two
volumes. The observables are normalized by the number of lattice points $V$.} 
\label{spincompare}
\end{figure}

As a first test of our numerical results we compare the outcome of the flux simulation at 
$\mu = 0$ to the results from a conventional simulation in the spin representation 
(which at $\mu = 0$ is possible). In Fig.~\ref{spincompare} we study 
on two different volumes the observables 
$U, C, \langle P \rangle$ and $\chi_P$, normalized by the number of sites $V$, 
as a function of the temperature $\tau$ at $\kappa = 0.005, \, \mu = 0$.  
  
\begin{figure}[t!]
\centering
\hspace*{8mm}
\includegraphics[width=0.90\textwidth,clip]{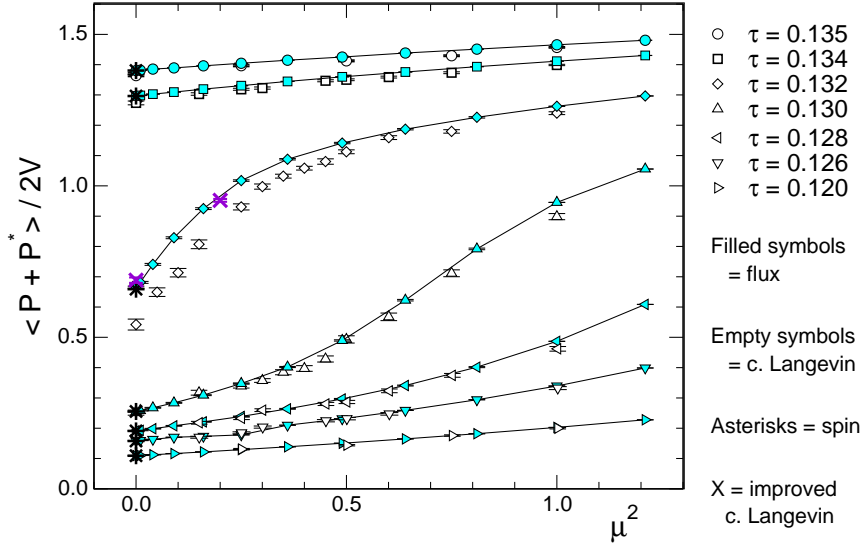}
\vspace*{-3mm}
\caption{Comparison of $\langle P + P^\star \rangle/2V$
 from the flux simulation (filled symbols) to the results from the complex Langevin approach 
(empty symbols and two high accuracy data points are marked with crosses). For $\mu = 0$ we also added the results from a simulation in the
conventional spin approach (asterisks). We compare data at different values of $\tau$ as a function of $\mu^2$ 
for $\kappa = 0.02$ on lattices of size $10^3$.} 
\label{langevincompare}
\end{figure}

For all four observables we find good agreement of the results from the flux simulation and
the data from the conventional approach. This confirms that the mapping from the
conventional representation to the flux degrees of freedom is correct, that the observables
are properly represented in the flux picture and that the flux Monte Carlo simulation 
works for $\mu = 0$. To check our approach and its implementation also at $\mu \neq 0$,
we compare our results also to the outcome of a perturbative $\tau$-expansion. This comparison 
will be discussed in the next section.

\begin{table}[b!]
\begin{center}
\begin{tabular}{l|cc|cc}
\hline
 & $\mu = 1.0$ & $\mu = 1.0$ & $\mu = 3.0$ & $\mu = 3.0$ \\ 
\hline
 & $\langle P \rangle$ & $\langle P^\star \rangle$ & 
  $\langle P \rangle$ & $\langle P^\star \rangle$ \\
\hline   
complex Langevin & 0.2419(19) & 0.3605(13) & 1.70615(27) & 1.74590(24)  \\
flux simulation  & 0.2416(11) & 0.3604(13) & 1.70627(14) & 1.74683(17)  \\
\hline
\end{tabular}
\end{center}
\caption{
Comparison to complex Langevin results with stepsize extrapolation  
\protect{\cite{aarts}} for $V = 10^3,\, \tau = 0.125, \, \kappa = 0.02$
and $\mu$ as listed in the table.}
\label{comparison}
\end{table}

Another interesting test of our approach is a comparison to the recently published results
from a complex Langevin simulation of the SU(3) spin model \cite{aarts} (for older results
with the complex Langevin approach see \cite{wyldkarsch,gausterer}). Such a comparison can be
done
also for non-zero chemical potential and thus tests our approach as a function of all three
couplings, $\tau, \kappa$ and $\mu$. In turn, an agreement of the complex Langevin and the
flux results is an important test also for the complex Langevin approach which currently sees a
lot of interesting development.

Fig.~\ref{langevincompare} shows that the flux results and the data from complex Langevin
agree very well for a wide range of parameters, and for $\mu = 0$ also with the results
from a conventional simulation. A slight discrepancy is seen for the $\tau = 0.132$ data when the
complex Langevin calculation is done
using the lowest-order Euler discretization at a single stepsize $\epsilon=0.0005$, 
without performing a zero-stepsize extrapolation. Since
observables in this discretization depend linearly on the stepsize,
the authors of \cite{aarts} provided us with improved data points for two values, $\mu^2 = 0.0$ and 0.2 at $\tau=0.132$,
where the higher-order algorithm discussed in \cite{aarts} was used, in which
stepsize corrections are essentially absent. These data points are represented by crosses  
in the plot and they nicely match the flux simulation results and for $\mu = 0$ also the 
data from the conventional approach.\footnote{We thank Gert Aarts and Frank James for providing us with the
data of \protect{\cite{aarts}} and extensive communication on their complex Langevin
simulation of the 
SU(3) spin model.}
    
\section{Phase diagram}

\begin{figure}[p]
\centering
\hspace*{-2mm}
\includegraphics[width=0.75\textwidth,clip]{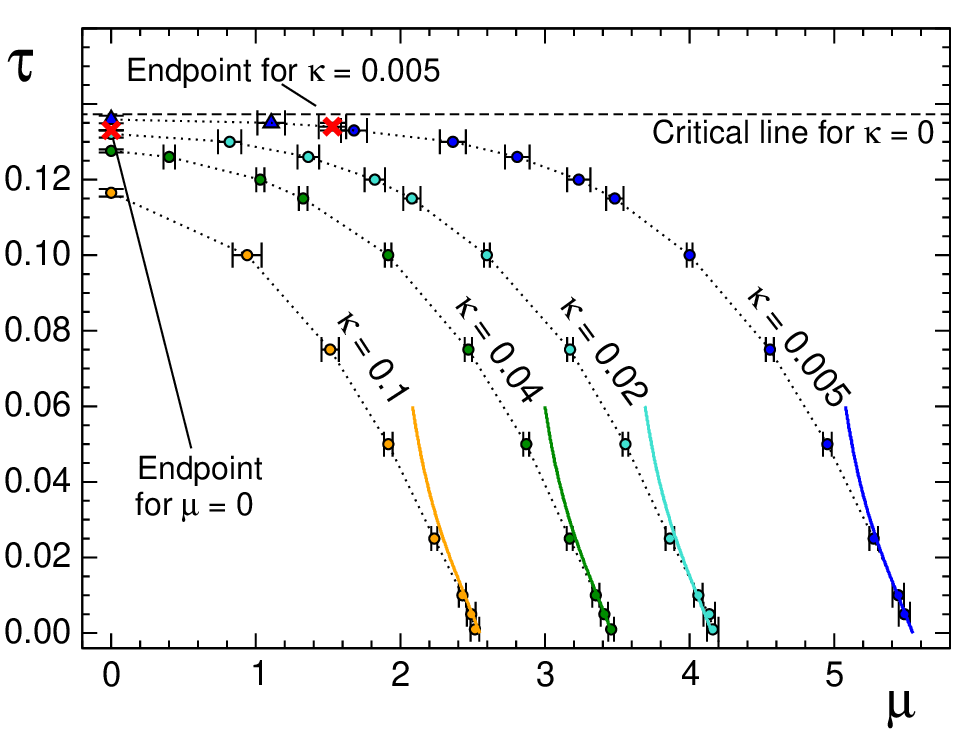}
\vspace*{-3mm}
\caption{The phase diagram as a function of $\tau$ and $\mu$ for different values of
$\kappa$. The phase boundaries (symbols connected with dotted lines)
were determined from the maxima of $\chi_P$ on $10^3$ lattices, or from a combined analysis
on $10^3, \, 16^3$ and $20^3$ lattices. The solid curves at small
$\tau$ are the results of an expansion in $\tau$. The horizontal dashed line  marks the
position of the critical $\tau$ for the $\kappa = 0$ case, and the two crosses indicate the
positions of the endpoints for the $\kappa = 0$ and $\kappa = 0.005$ cases (see
the text for details).} 
\label{phasediagram_Xp}
\vskip5mm
\centering
\hspace*{-2mm}
\includegraphics[width=0.75\textwidth,clip]{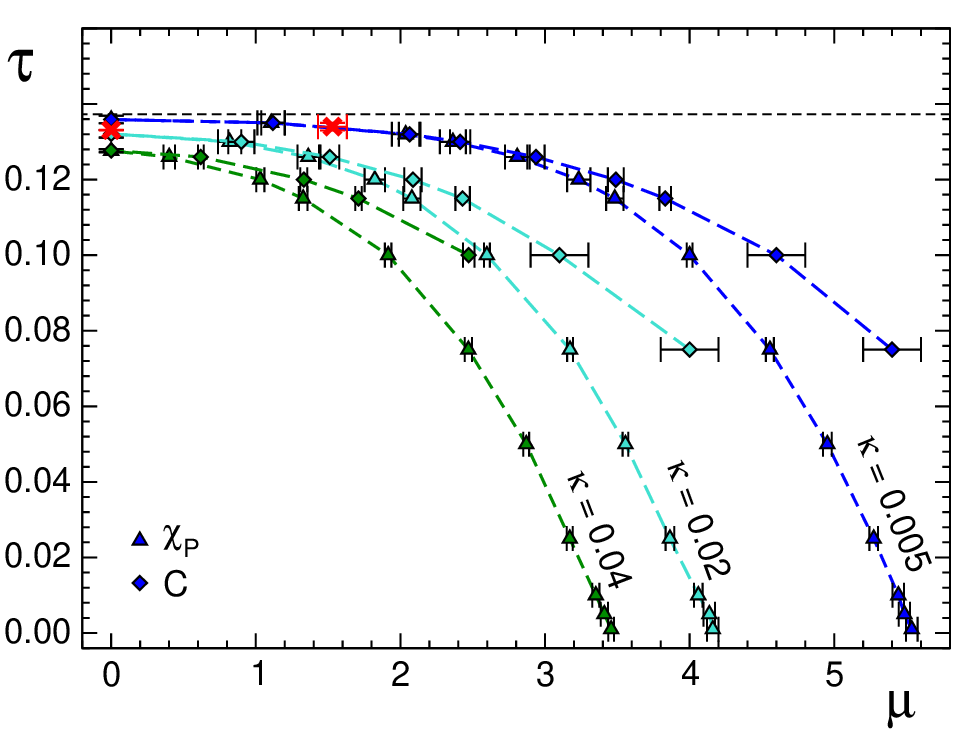}
\vspace*{-3mm}
\caption{Comparison of the maxima of $\chi_P$ (triangles) and $C$ (diamonds).} 
\label{phasediagram_compare}
\end{figure}

Let us now come to the analysis of the phase diagram as a function of $\tau$ and $\mu$. We
first present the results in  Figs.~\ref{phasediagram_Xp} and \ref{phasediagram_compare} and
discuss the details of its determination subsequently. In Fig.~\ref{phasediagram_Xp} we show
the positions of the maxima of $\chi_P$ in the $\tau$-$\mu$ plane 
(symbols connected by dotted
lines) for four
different values of $\kappa$. We used lattices of size $10^3$ or a combination of
$10^3, \, 16^3$ and $20^3$ lattices for the critical points and additional checks at
some of the parameter values. 
The solid curves at the bottom are the positions of the maxima from a perturbative
expansion in $\tau$ and obviously the Monte Carlo data nicely approach these results. This
again illustrates the correctness and accuracy of our simulation in the flux approach.

The horizontal dashed line indicates the critical value $\tau_c \sim 0.137$ of the theory
without external field, i.e., at $\kappa = 0$. At this temperature the $\kappa = 0$ model
undergoes a first order phase transition from a confining phase ($\langle P \rangle = 0$ for
$\tau < \tau_c$) into a deconfined phase with $\langle P \rangle \neq 0$. For $\mu = 0$ 
the first order transition persists for sufficiently small $\kappa$. Above some $\kappa_c$,
which defines a critical endpoint, only a crossover type of behaviour is observed. 
We estimated this endpoint to be at $\kappa_c = 0.016(2),\, \tau_c = 0.1331(1)$.
It is marked with a red cross in Fig.~\ref{phasediagram_Xp}. For values of 
$\kappa$ smaller than $\kappa_c$ a first order transition separates the confined and the
deconfining phase at some critical $\tau_c \in [0.133,\,0.137]$ which depends on $\kappa$.

When the chemical potential is turned on, the transition is weakened further, i.e., the 
position of the endpoint is shifted towards smaller values of $\kappa$. We determined the
endpoint for one other line in the phase diagram, the curve for $\kappa = 0.005$. For this
curve we find the endpoint to be at $\tau_c = 0.134(1),\, \mu_c = 1.53(10)$. Also this
endpoint is marked with a red cross in Fig.~\ref{phasediagram_Xp}. Points on the $\kappa =
0.005$ curve to the left of that endpoint correspond to a first order transition, 
while points to the right are characterized by crossover behavior.

In Fig.~\ref{phasediagram_compare} we compare the positions of the maxima of
$\chi_P$ (triangles) to the positions of the maxima of the heat capacity $C$ (diamonds). 
The comparison is done for three values of $\kappa$. 
For parameter values where we determined first order transitions (upper left corner) the
positions of the maxima agree for the two second derivatives of the free energy. However,
once we enter the region of crossover behavior without any singularities there is no
reason to expect that the maxima coincide. Indeed we observe this scenario: As one
increases $\mu$, the curves for the maxima of $\chi_P$ and $C$ start to
separate and the right half of the phase diagram has only a very broad crossover between
the $\langle P \rangle \sim 0$ region (left of the curves) and the $\langle P \rangle \neq 0$
phase.

\begin{figure}[p]
\centering
\hspace*{-4mm}
\includegraphics[width=1.05\textwidth,clip]{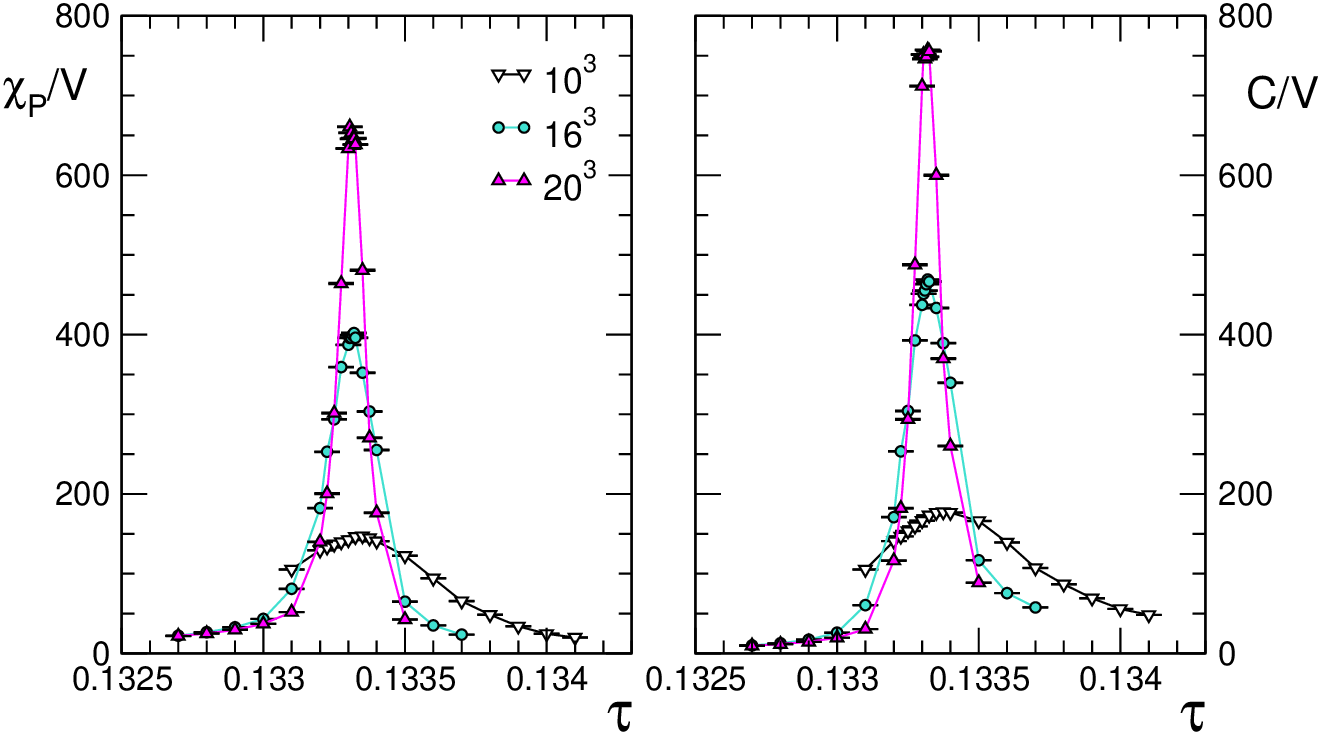}
\vspace*{-7mm}
\caption{Polyakov loop susceptibility and heat capacity as a function of $\tau$ 
for $\kappa = 0.015$ and $\mu = 0$. We compare three different volumes.} 
\label{susceptibilities}
\centering
\vskip10mm
\hspace*{-4mm}
\includegraphics[width=1.05\textwidth,clip]{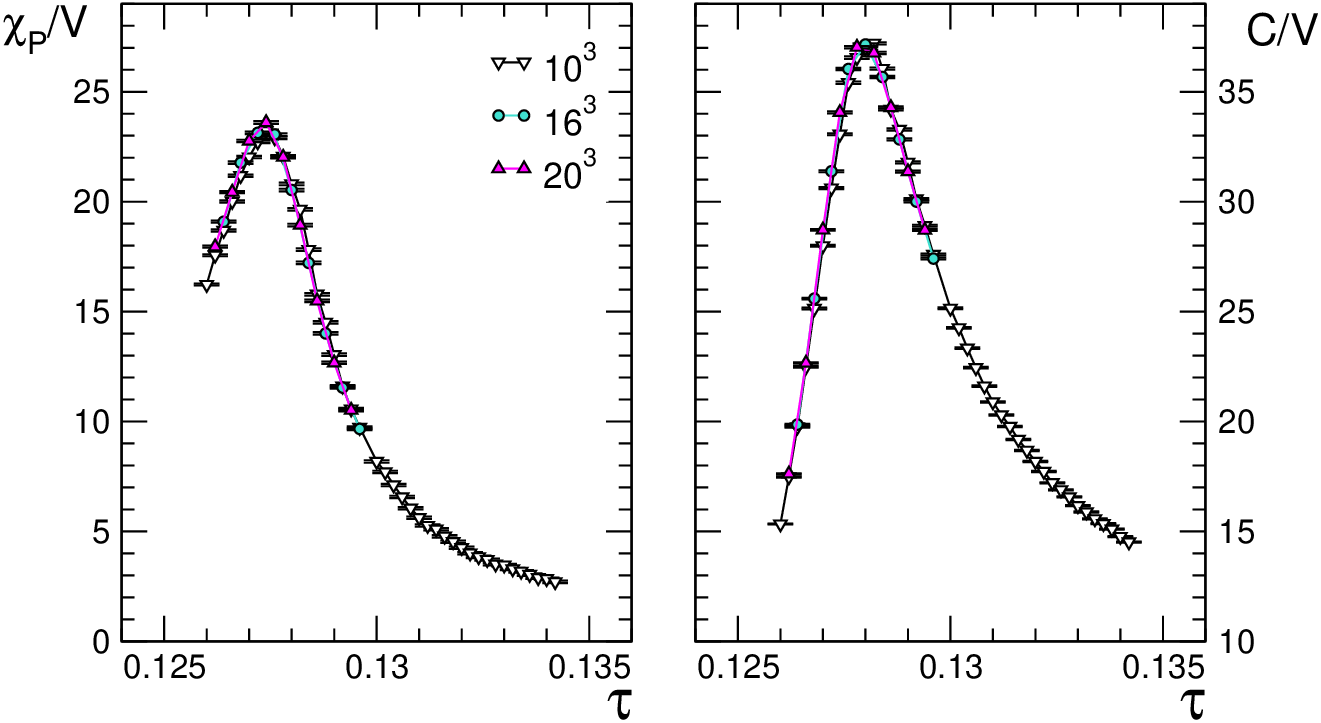}
\vspace*{-7mm}
\caption{Polyakov loop susceptibility and heat capacity as a function of $\tau$ 
for $\kappa = 0.040$ and $\mu = 0$. We compare three different volumes.} 
\label{susceptibilities2}
\end{figure}

Let us now analyze the behavior of our observables near the transition/crossover lines in
more detail. In Fig.~\ref{susceptibilities} we show the Polyakov loop
susceptibility $\chi_P$ and the heat capacity $C$ as a function of $\tau$ for 
$\kappa = 0.015$ and $\mu = 0$. This value of $\kappa$ is in the range where for
$\mu = 0$ one still observes a first order transition for some critical value
$\tau_c$. The figure shows that both, $\chi_P$ and $C$ develop a sharp peak which
scales with the volume. Furthermore, both these second derivatives of the free
energy peak at the same position, and we can read off $\tau_c \sim 0.1333$.

The situation is different when we look at a larger value of $\kappa$. 
In Fig.~\ref{susceptibilities2} we show again $\chi_P$ and 
$C$ as a function of $\tau$, but now at $\kappa = 0.04$ and $\mu = 0$. Although
the two quantities still show a maximum, it is obvious that the height of the
peak does not scale with the volume, and the curves for our three
volumes fall on top of each other. Thus we conclude that at $\kappa = 0.04$ we only observe a crossover
type of transition. We furthermore point out, that the positions of the maxima 
of $\chi_P$ and $C$ do not coincide, another indication for a crossover type of
behavior. 

\begin{figure}[p]
\centering
\includegraphics[width=1.0\textwidth,clip]{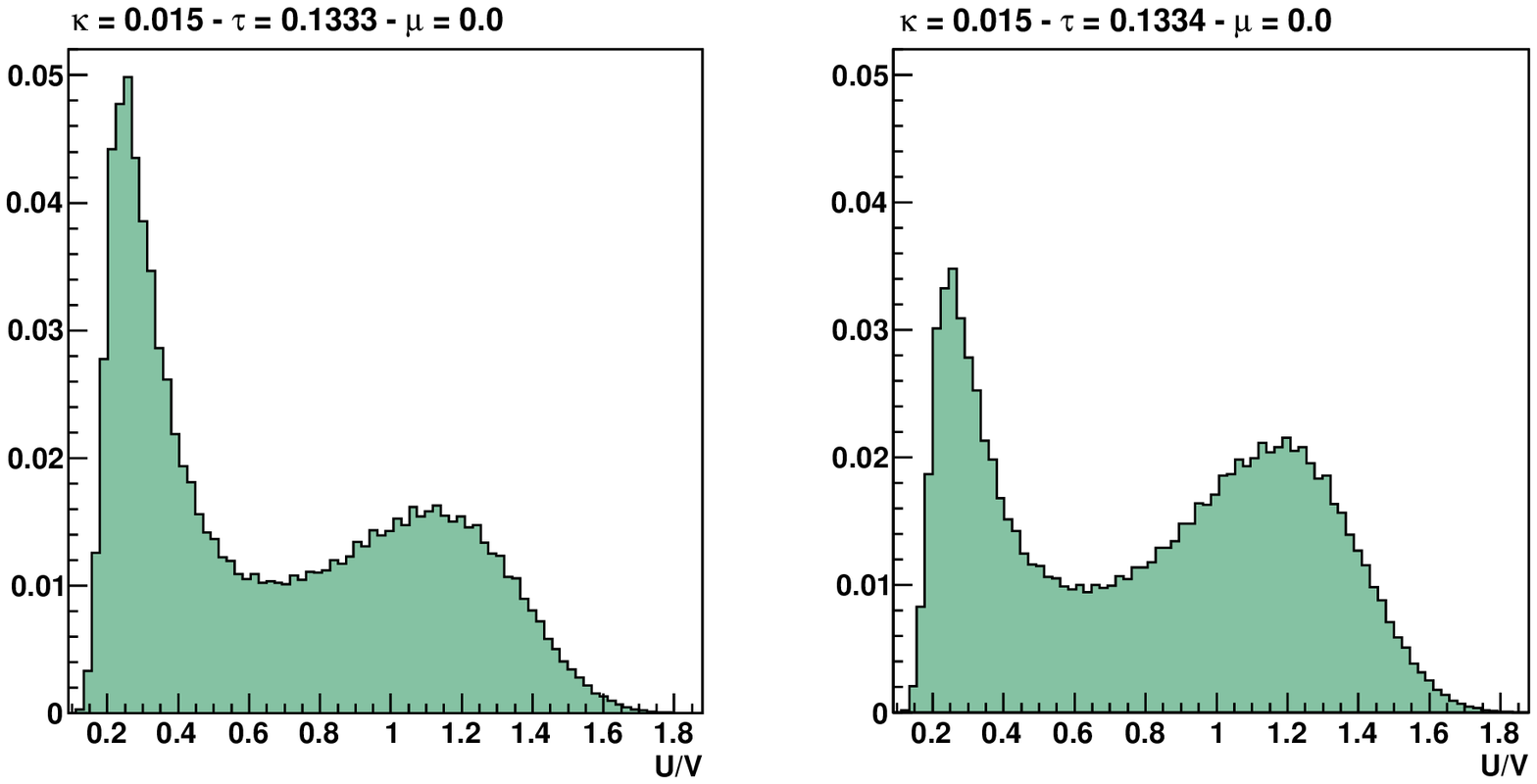}
\vspace*{-3mm}
\caption{Histograms for the distribution of the values for the internal energy
$U/V$ at $\kappa = 0.015$ and $\mu = 0$ for two different values of $\tau$ close to
the critical value.} 
\label{histogram1}
\vskip5mm
\centering
\includegraphics[width=1.0\textwidth,clip]{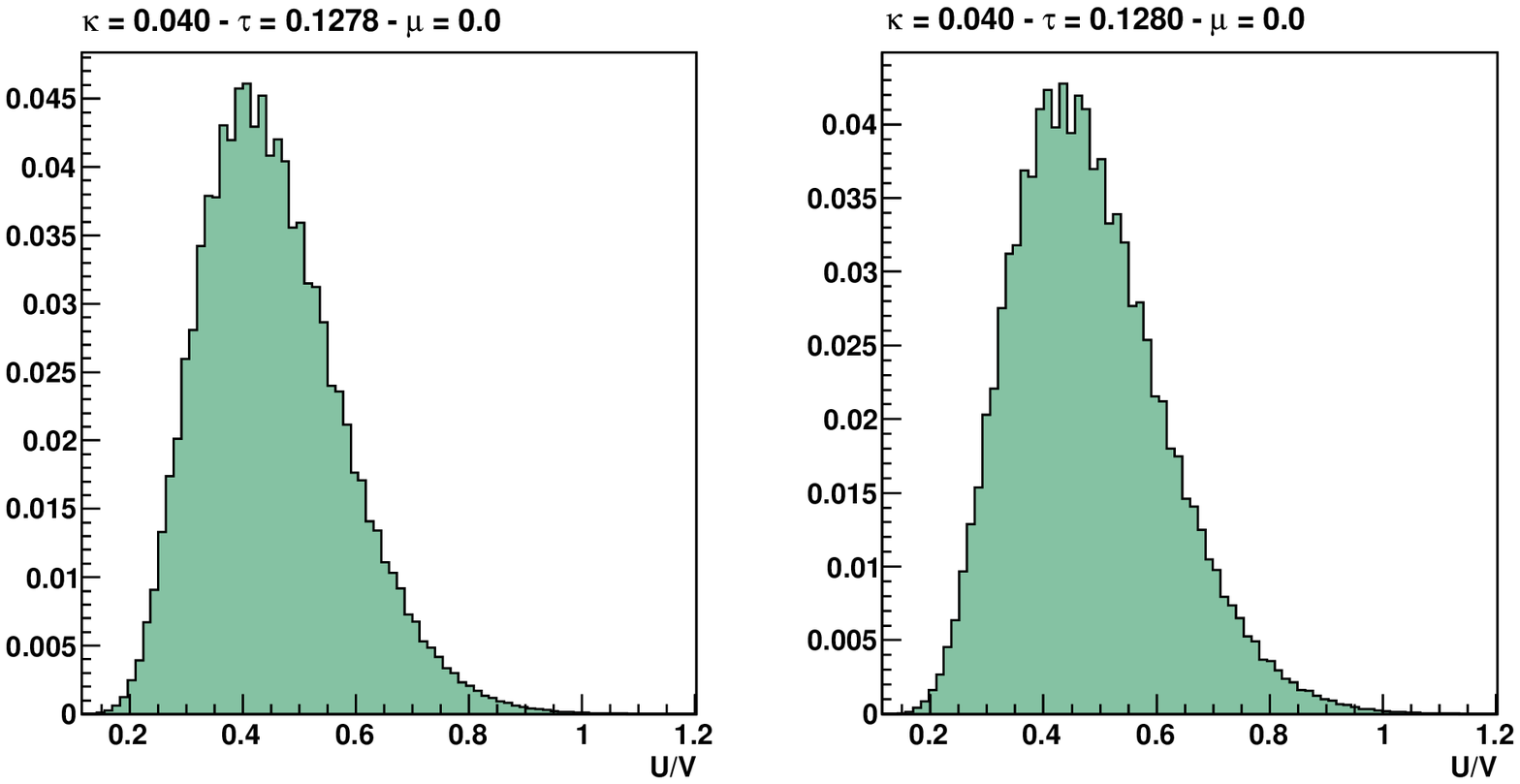}
\vspace*{-3mm}
\caption{Histograms for the distribution of the values for the internal energy
$U/V$ at $\kappa = 0.040$ and $\mu = 0$ for two different values of $\tau$ close 
to the position of the maximum of $C$.} 
\label{histogram2}
\end{figure}

We cross-checked the different behavior (first order versus crossover) at 
$\kappa = 0.015$ and $0.040$ by inspection of histograms for the distribution of
the internal energy $U$. In Fig.~\ref{histogram1} we show the distribution of
$U$ normalized by the volume, for the $\kappa = 0.015, \, \mu = 0$ case that
corresponds to Fig.~\ref{susceptibilities}. We use two values of $\tau$ close to
the critical value $\tau_c \sim 0.1333$ that we determined from the position of
the peaks of $\chi_P$ and $C$. The familiar two-state signal
that characterizes first order transitions is clearly established. 
The situation is different for $\kappa = 0.040, \; \mu = 0$
(Fig.~\ref{histogram2}). Here we observe only a single peak in the histogram
which changes smoothly with varying $\tau$, a behavior characteristic of a
crossover. 

Thus from the absence of volume scaling (Fig.~\ref{susceptibilities}) and the
single peak histogram (Fig.~\ref{histogram2}) we conclude that for $\kappa =
0.04$ we do not observe first order behavior. This finding is different from 
a mean field analysis of the SU(3) spin model \cite{karschMF} 
where a first order behavior was reported for
values of $\kappa$ up to a critical endpoint at $\kappa = 0.059$. Our results
show clearly that the endpoint must be below $\kappa = 0.04$ and our best
estimate is the value given already above in the discussion of
Fig.~\ref{phasediagram_Xp}, $\kappa_c = 0.016(2)$. 

\begin{figure}[t]
\centering
\hspace*{-4mm}
\includegraphics[width=1.05\textwidth,clip]{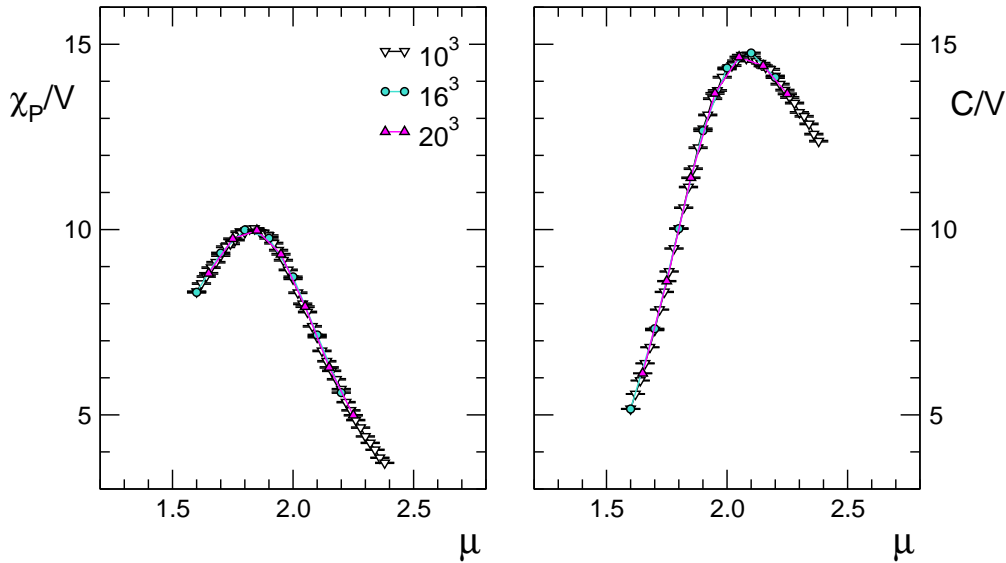}
\vspace*{-7mm}
\caption{Polyakov loop susceptibility and heat capacity as a function of $\mu$ 
for $\kappa = 0.02$ and $\tau = 0.12$. We compare three different volumes.} 
\label{susceptibilities3}
\end{figure}

We conclude our discussion of the phase diagram with an example at different 
values of the parameters. So far we presented thermodynamic quantities as a
function of the temperature parameter $\tau$. In Fig.~\ref{susceptibilities3} we
show again Polyakov loop susceptibility $\chi_P$ and heat capacity $C$, but now
as a function of chemical potential $\mu$ at fixed $\tau = 0.12$ and $\kappa =
0.02$. In other words, in Figs.~\ref{phasediagram_Xp} and
\ref{phasediagram_compare} this corresponds to a horizontal cut through the
$\kappa = 0.02$ crossover region at $\tau = 0.12$. From the fact that the
maxima of $\chi_P$ and $C$ do not coincide in that parameter region we have 
already concluded in the discussion of Fig.~\ref{phasediagram_compare} that this
is a crossover region. The absence of volume scaling for $\chi_P$ and $C$
confirms this conclusion. 

\section{Summary}

In this paper we have presented a Monte Carlo simulation of the SU(3) spin model
in a flux representation. Using the flux degrees of freedom the complex phase
problem is removed and a Monte Carlo simulation becomes possible also at finite
chemical potential. After a suitable reparametrization of the flux
representation \cite{gattringer} we presented a Monte Carlo algorithm that
alternates updates for the two kinds of monomer and dimer variables. 

Our approach was carefully tested: For $\mu = 0$ we compared our results 
to a Monte Carlo simulation using the conventional spin representation.
For small $\tau$ and arbitrary $\mu$ our Monte Carlo results perfectly match the
outcome of a perturbative expansion in $\tau$. These tests confirm the validity
of the map to the flux variables, the correct representation of the observables
and test the implementation of the algorithm. 

We also compared the recently published results from a complex Langevin simulation of the SU(3)
spin model \cite{aarts} to the data from our flux simulation.
We find very good agreement between the two methods which is a valuable test for
both, the flux and the complex Langevin approach.
 
Having tested and compared the implementation of the Monte Carlo method in the
flux approach we continued with a systematic analysis of the phase diagram in
the $\tau$-$\mu$ plane. The phase boundaries were determined from the maxima of 
the Polyakov loop susceptibility $\chi_P$ and the heat capacity $C$. We found that for small
$\kappa$ and $\mu$ the transition is of first order. In these cases $\chi_P$ and $C$ display
volume scaling, the histograms show a two-state signal and the positions of the maxima of  
$\chi_P$ and $C$ coincide. For larger values of $\kappa$ and $\mu$ we only found crossover
behavior. The crossover region widens with increasing $\mu$, i.e., the positions of the maxima
of $\chi_P$ and $C$ become more separated.

The analysis of the SU(3) spin model in this paper is the first complete mapping of the
phase diagram for a non-abelian
gauge group. We hope that the techniques developed here provide a useful step towards 
new approaches for the simulation of QCD or QCD-like systems with chemical potential.

\section*{Acknowledgments}
We thank Gert Aarts, Hans Gerd Evertz, Daniel G\"oschl, Frank James,
Christian Lang, Gundolf Haase and Manfred Liebmann
for fruitful discussions at various stages of this work, and
Gert Aarts and Frank James also for providing the complex Langevin data used in 
Fig.~2. Y.~Delgado
is supported by the FWF Doktoratskolleg {\sl Hadrons in Vacuum, Nuclei and
Stars} (DK W1203-N08) and by the Research Executive Agency (REA) of the European Union 
under Grant Agreement number PITN-GA-2009-238353 (ITN STRONGnet). C.~Gattringer thanks
the members of the INT at the Physics Institute of the University of Washington, Seattle, 
where part of
this work was done, for their hospitality and the inspiring atmosphere. C.~Gattringer also
acknowledges 
support by the Dr.\ Heinrich J\"org Stiftung, Karl-Franzens-Universit\"at Graz, Austria.

\end{document}